\documentclass[aps,pra,onecolumn,12pt,tightenlines,superscriptaddress,notitlepage,nofootinbib]{revtex4-1}

\newcommand{\eqnref}[1]{Eq.~(\ref{#1})}

\newcommand{\refapp}[1]{Appendix \ref{#1}}

\newcommand{\ket}[1]{|#1\rangle}

\usepackage{amsmath}
\usepackage{amsthm}
\usepackage{amsfonts}
\usepackage{graphicx}
\usepackage{fullpage}
\usepackage{comment}
\usepackage{color}
\usepackage{hyperref}
\usepackage{mathtools}
\usepackage{todonotes}
\usepackage{pifont}
\usepackage{chngpage}

\newcommand{\cmark}{\ding{51}}%
\newcommand{\xmark}{\ding{55}}%

\newtheorem{lemma}{Lemma}
\newtheorem{thm}{Theorem}
\newtheorem{definition}{Definition}

\begin{document}
\title{An improved Lieb-Robinson bound for many-body Hamiltonians with power-law interactions}
\author{Dominic V. Else}
\affiliation{Department of Physics, Massachusetts Institute of Technology, Cambridge, MA 02139, USA}
\affiliation{Department of Physics, University of California, Santa Barbara, CA 93106, USA}
\author{Francisco Machado}
\affiliation{Department of Physics, University of California, Berkeley, CA 94720, USA}
\author{Chetan Nayak}
\affiliation{Microsoft Research, Station Q, University of California, Santa Barbara, CA 93106, USA}
\affiliation{Department of Physics, University of California, Santa Barbara, CA 93106, USA}
\author{Norman Y. Yao}
\affiliation{Department of Physics, University of California, Berkeley, CA 94720, USA}
\affiliation{Materials Sciences Division, Lawrence Berkeley National Laboratory, Berkeley CA 94720, USA}

\begin{abstract}
In this work, we prove a new family of Lieb-Robinson bounds for lattice spin systems with long-range interactions.
Our results apply for arbitrary $k$-body interactions, so long as they decay with a power-law greater than $kd$, where $d$ is the dimension of the system. 
More precisely, we require that the sum of the norm of terms with diameter greater than or equal to $R$, acting on any one site, decays as a power-law $1/R^\alpha$, with $\alpha > d$.
These new bounds allow us to prove that, at any fixed time, the spatial decay of quantum information follows arbitrarily closely to $1/r^{\alpha}$.
Moreover, we define a new light-cone for power-law interacting quantum systems, which captures the region of the system where changing the Hamiltonian can affect the evolution of a local operator.
In short-range interacting systems, this light-cone agrees with the conventional definition. 
However, in long-range interacting systems, our definition yields a stricter light-cone, which is more relevant in most physical contexts. 
\end{abstract}

\maketitle

In a relativistic quantum field theory, information can never travel faster than the speed of light. A \emph{Lieb-Robinson bound} 
\cite{Lieb__72,Nachtergaele_0506,Hastings_0507,Nachtergaele_0712_3318,Nachtergaele_0712_3820,Nachtergaele_1004}
 establishes a similar ``light-cone'' for the spread of quantum information in a non-relativistic lattice system.
However, the information spread outside the light-cone is not strictly
vanishing but, instead, has non-zero tails.
Such constraints on the spread of information, in addition to being physically important in their own right, have also been used as ingredients in the rigorous mathematical proof of key results about non-relativistic lattice systems \cite{Hastings_0305,Nachtergaele_0506, Hastings_0503,Hastings_0507,Osborne_0601,Nachtergaele_0608,Bravyi_0603,Hastings_0705,Bravyi_1001_0344,Bravyi_1001_4363,Bachmann_1102,Michalakis_1109,Abanin_1509}, including the exponential decay of correlations in the ground states of gapped Hamiltonians \cite{Nachtergaele_0506,Hastings_0507}
and the stability of topological order \cite{Bravyi_0603,Bravyi_1001_0344,Bravyi_1001_4363,Michalakis_1109}.

More recently, new results have generalized Lieb-Robinson bounds to
lattice spin systems with interactions between distant spins that fall off as a
power of their separation \cite{Hastings_0507,Hauke_1304,Eisert_1309,Richerme_1401,FossFeig_1410,Matsuta_1604}. 
Such long-range interactions arise in a wide variety of experimental platforms, ranging from 
solid-state spin defects \cite{Ruderman__54,Kasuya__56,Yosida__57} to quantum optical systems of trapped ions \cite{Blatt__12}, polar molecules \cite{Moses_1610}, and Rydberg atoms \cite{Saffman_0909}.
While the majority of previous studies have focused on few-body physics, recent advances have enabled a number of these platforms to begin probing the many-body dynamics and information propagation of strongly interacting, long-range systems \cite{Yan_1305,Richerme_1401,Zeiher_1705}.

Motivated by the development of these physical platforms, in this work, we improve Lieb-Robinson bounds for generic power-law interactions.  
Specifically, let us consider a system of spins on a lattice $\Lambda$ governed by a Hamiltonian $H$, which can be written as a sum $H = \sum_Z H_Z$ of terms acting on sets of lattice sites $Z\subseteq\Lambda$ in $d$-dimensional space.  Moreover, we assume (among other conditions described in Section \ref{sec:assumptions}) that there exists a constant $J$ such that
\begin{equation}
\sup_{z \in \Lambda} \sum_{Z \ni z: \mathrm{diam}(Z) \geq R}
||{H_Z}||\,\,
\leq \,\,\frac{J}{R^{\alpha}},
\end{equation}
where $\mathrm{diam}(Z)$ is the greatest distance between any two points in $Z$.
A familiar example \cite{Zhang_1609, Choi_1610} is the long-range Ising interaction,
\begin{equation}
H = H_{\text{short-range}} + J\sum_{i\neq j} \frac{1}{|\textbf{r}_i - \textbf{r}_j|^{d+\alpha}} \sigma_i^z \sigma_j^z.
\end{equation}

An early result on Lieb-Robinson bounds in power-law interacting systems was proved in Ref.~\cite{Hastings_0507}, which demonstrated the existence of a
light-cone whose size grows exponentially in time for any $\alpha >
0$.
More recently, this result was improved in
Refs.~\cite{FossFeig_1410} and \cite{Matsuta_1604},
where it was shown that a \emph{power-law} light-cone emerges for $\alpha > d$, where $d$ is the spatial dimension.

However, each of these results has certain limitations (Table~I).
On the one hand, Ref.~\cite{FossFeig_1410} assumes a two-body Hamiltonian,
where each term acts on at most two spins.~\footnote{Note that we are following the convention of Ref.~\cite{Matsuta_1604} in the definition of the power-law, $\alpha$. 
This differs from the definition of  ``$\alpha$'' in Ref.~\cite{FossFeig_1410}, which would be equal to $\alpha+d$ in our convention.}
This assumption limits the usage of this result in analyzing multi-body effective Hamiltonians
of  broad interest in condensed matter physics.
Such Hamiltonians can arise in a number of different contexts:
for example, ring-exchange interactions may be important in solid $^3$He \cite{Thouless__65} and are known to stabilize certain topological phases \cite{Misguich_9812,Paramekanti_0203};
multi-body Hamiltonians arise in explicit constructions of various results in mathematical physics \cite{Hastings_0503,Bravyi_1001_0344,Bravyi_1001_4363, Abanin_1509};
and higher-body interactions naturally emerge in the effective description of periodically-driven two-body Hamiltonians \cite{Bukov_1407, Abanin_1509}.

\begin{table}
\begin{tabular}{|c|c|c|c|c|} \hline
  Reference & Multi-body Hamiltonians & Asymptotic Spatial Decay & LC1 &  LC2  \\ \hline
  Foss-Feig et al.~Ref.~\cite{FossFeig_1410} & \xmark & $r^{-(\alpha + d)}$ & $\alpha > d$ & $\alpha > d$ \\
  Matsuta et al.~Ref.~\cite{Matsuta_1604} & \cmark & $r^{ -(\alpha-d)/(\eta +1)}$& $\alpha > d$ & $\alpha > 2d$\\
  Present work & \cmark & $r^{-\alpha}$ & $\alpha > d$ & $\alpha > d$\\
 \hline
\end{tabular}
\caption{Summary of power-law Lieb-Robinson bounds for $\alpha > d$. Note that the LC1 and LC2 columns describe the power-law regime where these light-cones exist and are power-law.}
\end{table}

On the other hand, while Ref.~\cite{Matsuta_1604} overcomes this two-body assumption, it proves a significantly weaker result regarding the power-law decay of  information outside the light-cone (Table I).\footnote{The $\sim 1/r^{(\alpha-d)/(\eta+1)}$ decay can be improved arbitrarily close to $\sim 1/r^{\alpha-d}$ at the cost of widening the light-cone.}
%
%
In particular, for $\alpha \gtrsim d$, the
bounds of Ref.~\cite{Matsuta_1604}  ensure only a relatively slow decay outside the light-cone;
%
which can limit its applicability to some important results, e.g. bounding the difference in operators time evolved under slightly different Hamiltonians.

In this paper, we prove a Lieb-Robinson bound that addresses both of the above concerns.
We demonstrate that  for multi-body interactions with $\alpha > d$, the spatial decay of quantum information outside the light-cone scales  arbitrarily closely to $\sim 1/r^{\alpha}$  (Table I).
While this bound is not as strong as the $\sim 1/r^{\alpha + d}$ decay obtained in Ref.~\cite{FossFeig_1410}, our combination of an improved scaling (over Ref.~\cite{Matsuta_1604}) and applicability to arbitrary multi-body Hamiltonians, enables the usage of this Lieb-Robinson bound to prove new results in mathematical physics \cite{Machado_18xx}.

An important comment is in order. Unlike either short-range or exponentially decaying interactions, power-law interactions are characterized by Lieb-Robinson bounds with power-law tails which lack a natural notion of a length scale. 
This implies that one should be particularly careful when defining an associated light-cone for such long-range interacting systems. 
One possible definition of a light-cone (used in Refs.~\cite{FossFeig_1410,Matsuta_1604}) is the following: at late times,  outside the light-cone, the propagated quantum information is small. From here on, we will refer to this as light-cone 1 (LC1). 
For short-range interacting systems, LC1 is the only length-scale associated with time evolution. 
For power-law interacting systems, one can already get a sense of the insufficiency of LC1 by noting the following: despite the fact that the asymptotic spatial decays obtained in Refs.~\cite{FossFeig_1410}, \cite{Matsuta_1604} and this work are quite different (Table I), they all yield the same LC1 (Table II).

To this end, we introduce a second light-cone, LC2, which properly captures these differences.
In particular, LC2 ensures that at late times, the evolution of the operator is not affected by changes to the Hamiltonian outside of LC2. 
For short-range interacting systems, LC1 and LC2 coincide, but for long-range interacting systems, they can be quite different.
More specifically, Ref.~\cite{FossFeig_1410} exhibits a finite, power-law LC2 for $\alpha > d$, while Ref.~\cite{Matsuta_1604} does not have a finite LC2 for $\alpha < 2d$, despite both having the same power-law LC1.
Intuitively, the lack of an LC2 for $\alpha > d$ in Ref.~\cite{Matsuta_1604} is due to the aforementioned slow asymptotic spatial decay of quantum information.
This highlights the importance of our improved decay; it enables us to prove our second main result, which is the existence  of a power-law LC2 for $\alpha > d$ for arbitrary multi-body Hamiltonians (Table I)\footnote{We note that this extension comes at the expense of a worse LC2 exponent compared to Ref.~\cite{FossFeig_1410} (Table II)}.



%

\section{Assumptions and notation}
\label{sec:assumptions}
Our notation will be similar to that of Ref.~\cite{Matsuta_1604}.
We consider a set of lattice sites $\Lambda$ with a metric $d(x,y)$ for $x,y \in \Lambda$, and a Hamiltonian $H$ written as a sum of terms $H = \sum_Z H_Z$, where $H_Z$ is supported on the set $Z \subseteq \Lambda$.
We extend the notation of the metric to sets, denoting $d(X,Y)$ as the minimum distance between any two elements of the sets $X,Y\subseteq \Lambda$, as well as between sets and sites, denoting $d(X,y) = d(X, \{y\})$.
We define a function $f(R)$
that captures the power-law decay of interactions:
\begin{equation}
\label{eqn:f(R)-def}
f(R) := \sup_{z \in \Lambda} \sum_{Z \ni z : \mathrm{diam}(Z) \ge R} \| H_Z \|~,
\end{equation}
where
\begin{equation}
\mathrm{diam}(Z) = \sup_{x,y \in Z} d(x,y)~,
\end{equation}
and we assume there are constants $J$ and $\alpha > d$ (the dimensionality of the system) such that $f(R) \leq J R^{-\alpha}$. We also require that the sum of the operator norms
of all of the terms involving any site be finite:
\begin{align}
\mathcal{C}_0 := \sup_{x \in \Lambda}\sum_{y \in \Lambda} \sum_{Z \ni x,y} \|H_Z\| < \infty~.
\end{align}

Finally, we assume certain regularity conditions on the lattice $\Lambda$. Specifically, we assume that $\Lambda$ can be embedded in Euclidean space $\mathbb{R}^d$, so that for each $z \in \Lambda$ there is a corresponding $\textbf{r}_z \in \mathbb{R}^d$, such that $d(x,y) = |\mathbf{r}_x - \mathbf{r}_y|$. Moreover, we assume there is a smallest lattice separation $a$ such that $d(x,y) \geq a$ for any $x,y \in \Lambda$ unless $x=y$. We choose to work in units such that $a=1$.

Let us also define $\tau_t^H(O)$ as the operator $O$ time-evolved according to
the Heisenberg representation
\begin{equation}
\tau_t^H(O) = e^{itH}\, O \, e^{-itH}.
\end{equation}

Throughout the paper we will use ``$C$'' to refer to any constants that depends only on $\sigma$ (the parameter introduced in the statement of the theorem in the next section) and the lattice.
It will not necessarily be the same constant each time it appears.


\section{Main result}

\begin{thm}
  \label{thm:thethm}
  Given the assumptions stated in Section \ref{sec:assumptions}, let observables $A$ and $B$ be supported on sets $X$ and $Y$ respectively. Then for any $(d+1)/(\alpha+1) < \sigma < 1$:
\begin{equation} \label{eqn:bound}
\| [ \tau_t^{H}(A), B ] \| \leq \|A\|\|B\| \left\{2|X| e^{vt - r^{1-\sigma}} + C_1\frac{\mathfrak{G}(vt)}{r^{\sigma \alpha}}\right\},
\end{equation}
where $r = d(X,Y)$ and $v = C_2 \max \{ J, \mathcal{C}_0 \}$. 
Moreover, there exists a constant $C_3$ such that:
\begin{equation}
  \mathfrak{G}(\tau) \leq C_3(\tau + \tau^{1+d/(1-\sigma)}) |X|^{n^*+2},
\end{equation}
where
\begin{align} \label{eq:nstar_def}
  n^* = \left\lceil \frac{\sigma d}{\sigma \alpha - d} \right\rceil
\end{align}
Here, all $C_i$ are constants only dependent on $\sigma$ and the lattice.
\end{thm}
By choosing $\sigma$ arbitrarily close to $1$ we obtain a decay of the Lieb-Robinson bound that approaches $\sim r^{-\alpha}$ for large $r$.


\section{Proof}

\subsection{Iteration Procedure}

The main challenge in understanding the spread of quantum information in long-range interacting systems is being able to differentiate the contribution from strong ``short'' range terms and the weak ``long'' range terms in a problem with no natural length scale.
As a result, there is no single separation between ``short'' and ``long'' range terms of the Hamiltonian that yields a strict bound.
We develop a construction that iteratively introduces a new short scale, enabling us to improve on the Lieb-Robinson bound by better accounting for the spatial decay of interactions in the Hamiltonian.

%
%
As a starting point, we consider a truncated version of our long-range Hamiltionian with a cutoff $R$, $H^{\le R}$:
\begin{align}
  H^{\le R} \,\, &= \sum_{Z : \mathrm{diam}(Z) \leq R} H_Z~.
\end{align}
At the end of our construction we can make  $R \to \infty$, recovering the full Hamiltonian.
Because $H^{\le R}$ has finite range $R$, a Lieb-Robinson bound for short-ranged Hamiltonians
can be applied.
However, this is clearly not the optimal bound,
as it assumes all interactions of range up to $R$ are equally strong, ignoring their decay with range.
Nevertheless, this provides the starting point for our iterative process.


At each iteration, the Hamiltonian $H^{\le R}$ is split into a short and a long-range piece using a cutoff $R'$:
\begin{align}
  H^{\le R} = H^{\le R'} + H^{R';R} \quad   \text{where} \quad H^{R';R} =  \sum_{Z : R'< \mathrm{diam}(Z) \le R} H_Z.
\end{align}
Then, following the strategy of Refs.~\cite{FossFeig_1410} and \cite{Matsuta_1604},
the time-evolution of an operator $A$ is separated into a contribution from the short-range part $H^{\le R'}$
and the long-range part $H^{R';R}$; these play the role,
respectively, of the unperturbed Hamiltonian
and the perturbation in the interaction representation.

Then, using Lemma 3.1 in Ref.~\cite{Matsuta_1604}, we can bound the total spread of the operator as a contribution from the short-range part $H^{\le R'}$, as well as an \emph{additional} contribution due to the long-range part $H^{R';R}$:
\begin{equation}
  \label{iterativecommutators}
 \| [\tau_t^{H^{\le R}}(A), B] \| \leq \| [\tau_t^{H^{\le R'}}(A), B ] \| +
2\| B \| \int_0^t \| [\tau_{t-s}^{H^{\le R'}}(A), H^{R';R}] \| ~ ds
\end{equation}
This procedure enables us to better distinguish the contribution of the strong short-range terms and the weak long-range terms of the evolution, improving upon the initial naive bound.
Once an improvement is obtained, one can perform the procedure again further reducing the contribution from the long-range piece of \eqnref{iterativecommutators} and improving the spatial decay of the Lieb-Robinson bound.
We note this iterative process recovers the argument of Ref.~\cite{Matsuta_1604} after one iteration; by iterating multiple times we can improve on their results.
We make this iterative construction more precise with the following Lemma:
\begin{lemma}
  \label{lem:iterative}
  Fix a set $X \subseteq \Lambda$ and a time $t$. Suppose that we have a function $\lambda^{(R)}(r)$ such that for all $0 \leq s \leq t$, $Y\subseteq\Lambda$ and observables $A$ and $B$ supported on sets $X$ and $Y$ respectively, the bound
\begin{equation}
\label{thebound}
\| [\tau_s^{H^{\le R}}(A), B] \| \leq \lambda^{(R)}(d(X,Y)) \| A \| \| B \|
\end{equation}
is satisfied. We assume that $\lambda^{(R)}(r)$ is monotonically increasing in $R$ and decreasing in $r$.
Then, for any $R' > 0$, \eqnref{thebound} is also satisfied with $\lambda$ replaced by $\widetilde{\lambda}$, defined according to
\begin{equation}
\label{eqn:lemma-recursion}
\widetilde{\lambda}^{(R)}(r) = \lambda^{(R')}(r) + C \Theta(R - R')\, |X|\, t\, f(R') \, \mathcal{I}[\lambda^{(R')}],
\end{equation}
where $f(R)$ is given in Eq. (\ref{eqn:f(R)-def}); $C$ is a constant independent of $R,R',|X|$ and $t$; $\Theta(x)$ is the Heaviside theta function and:
\begin{equation}
\label{eqn:recursion-integral}
\mathcal{I}[\lambda] = \lambda(0) + \int_{1/2}^{\infty} \rho^{d-1} \lambda(\rho) ~d\rho.
\end{equation}
\begin{proof}For $R' \geq R$, the result follows directly from the monotonicity
  with respect to $R$. On the other hand, for $R' < R$ we have, from  \eqnref{iterativecommutators}:
\begin{align}
 \| [\tau_t^{H^{\le R}}(A), B] \| &\leq \| [\tau_t^{H^{\le R'}}(A), B] \| + 2\| B \| 
\int_0^t \| [\tau_{t-s}^{H^{\le R'}}(A), H^{R';R} ] \| \, ds
 \cr
 &\leq \| [\tau_t^{H^{\le R'}}(A), B] \|  + 2\| B \|\int_0^t \sum_{Z:R' < \mathrm{diam}(Z) \le R} \| \bigl[ \tau_{s}^{H^{\le R'}}(A),H_Z \bigr] \| \, ds \cr
 &\leq \lambda^{(R')}(d(X,Y))
\| A \| \| B \|
+ 2t \| B \|  \!\!\!\! \sum_{Z:R' < \mathrm{diam}(Z) \le R} \lambda^{(R')}(d(X,Z)) \| H_Z \| \| A \| \cr
  &\leq \lambda^{(R')}(d(X,Y)) \| A \| \| B \|
+ 2t \| A \|  \| B \|  \sum_{z \in \Lambda} ~\sum_{Z \ni z: R' < \mathrm{diam}(Z) \le R} \lambda^{(R')}(d(X,z))  \| H_Z \| \cr
  &\leq \lambda^{(R')}(d(X,Y)) \| A \| \| B \| + 2t \| A \| \| B \|
 f(R') \sum_{z \in \Lambda} \lambda^{(R')}(d(X,z)) \cr
  &\leq \| A \| \| B \|\left\{\lambda^{(R')}(d(X,Y)) +2 t  f(R') |X| \sup_{x \in X} \sum_{z \in \Lambda} \lambda^{(R')}(d(x,z))\right\}.\cr
\label{eqn:recursion}
 \end{align}
In going from the second line to the third line, it is helpful to recall that
$\lambda^{(R')}(d(X,Y))$ is independent of $s$ (but dependent on $t$).
In going from the fourth line to the fifth, we used:
\begin{align}
\sum_{z \in \Lambda} \sum_{Z \ni z, R' < \mathrm{diam}(Z) < R} \lambda^{(R')}(d(X,z))  \| H_Z \| 
&=
\sum_{z \in \Lambda}\lambda^{(R')}(d(X,z))  \sum_{Z \ni z, R' < \mathrm{diam}(Z) < R}  \| H_Z \|
\cr
&\leq f(R') \sum_{z \in \Lambda}\lambda^{(R')}(d(X,z)) .
\end{align}
To get the final result, we use Lemma \ref{lem:sumtointegral} from the Appendix to replace the sum by an integral in the last line of Eq. (\ref{eqn:recursion}).

Finally, let us note that the simplest bound for $\lambda^{(R')}(d(X,Y))$ corresponds to the short-range Lieb-Robinson bound where the interactions have at most range $R'$.
 \end{proof}
\end{lemma}

\begin{figure}[t]
  \centerline{
    \includegraphics[width=6in]{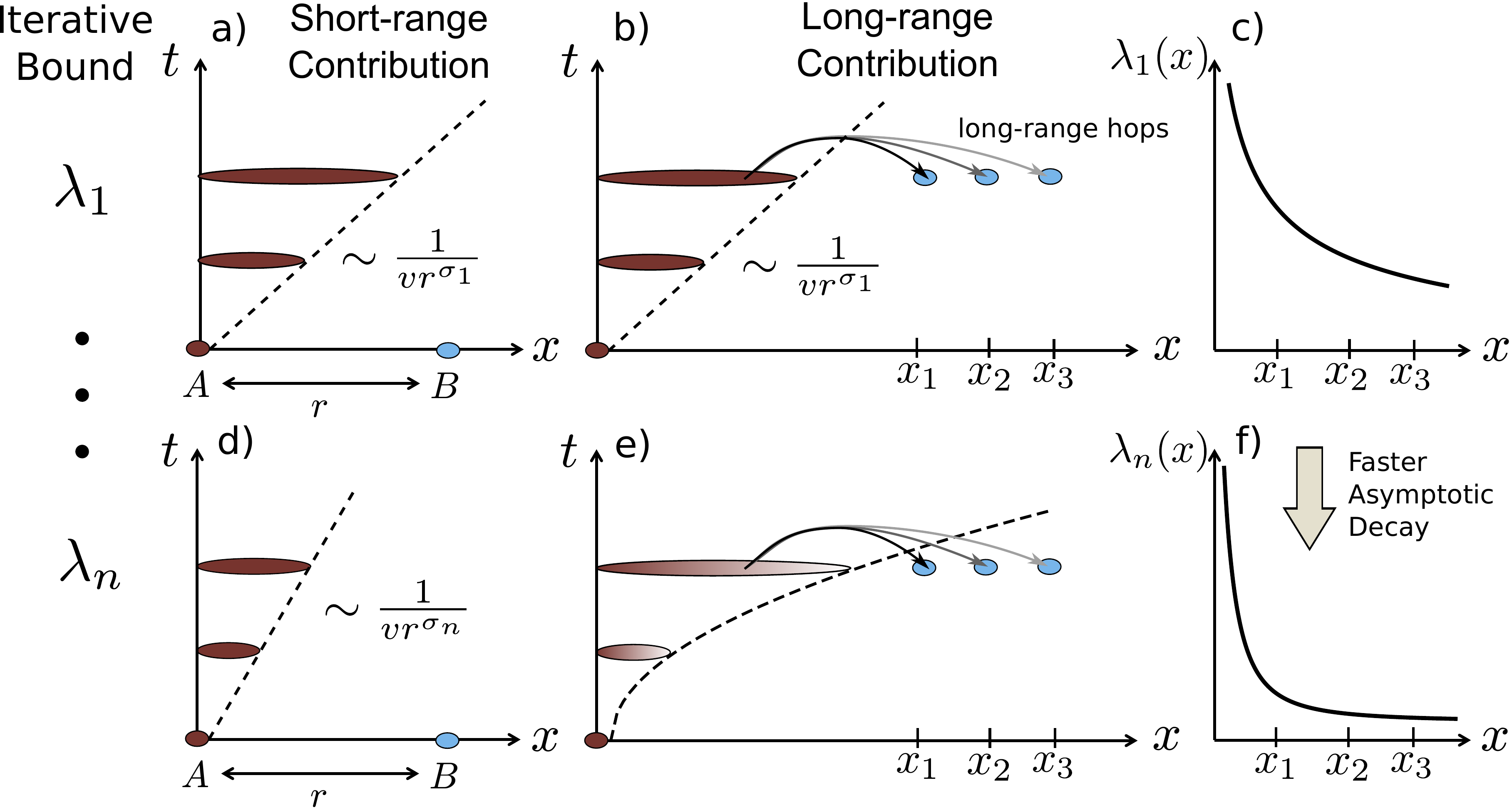}}
  \caption{
    The Lieb-Robinson bound measures the spread of quantum information during evolution by bounding the commutator of a time evolved local operator $A$, with another operator $B$ a distance $r$ away.
    The spread the operator $A$ can be apportioned into the spread due
    to interactions of range shorter than $R'$ (left column) and long-range hops due to interactions
    of range larger than $R'$ (center column).
    The long-range hops connect the short-range time evolved operator $A$ with strength
    at most $f(R')$ but they can originate from any location that $A$ has spread to,
    so the total contribution
    of these long-range hops is weighted by the integral $\mathcal{I}[\lambda]$ (see Lemma \ref{lem:iterative}).
    At the first iterative step, which yields $\lambda_1$, the short-range interactions can always be characterized by an exponentially decaying bound with a sharp light-cone with slope $(vR')^{-1}$, \eqnref{eq:short_Lieb}.
    This corresponds exactly to the short-range contribution to $\lambda_1$ ({\bf a}).
    The long-range contribution arises from the long-range hops that connect the inside of the light-cone to the support of $B$ ({\bf b}).
    By choosing the cut-off $R'$ as a function of the operator distance, $R'=r^{\sigma_1}$, the resulting bound becomes the sum of exponential and power-law decaying terms, \eqnref{lambda1}, the latter of which dominate the long distance decay of the bound ({\bf c}).
    This choice of $R'$ leads to the light-cone slope of $(v r^{\sigma_1})^{-1}$ of panel {\bf (a)}.
    At the $n$-th iteration step, which yields $\lambda_n$, we choose a new cut-off $\tilde{R}'$.
    As before, we obtain a short-range contribution that yields a linear light-cone with slope $(v\tilde{R}')^{-1}$ ({\bf d}).
    More importantly, the long-range hops will now be weighted by the power-law decay of the previous bound $\lambda_{n-1}$, illustrated by the red shading ({\bf e}).
    It is the combination of these two power-law decays that enables our iterative procedure to improve the asymptotic decay of the bound $\lambda_n$ after specifying the cut-off as $\tilde{R}' = r^{\sigma_n}$ ({\bf f}) (see Section III.B).
    This choice of $\tilde{R}'$ leads to the light-cone slope of $(v r^{\sigma_n})^{-1}$ of panel {\bf (d)}.
%
%
%
}
  \label{fig:iteration}
\end{figure}

We now iteratively apply Lemma \ref{lem:iterative}.
\eqnref{eqn:lemma-recursion} says that a Lieb-Robinson bound $\lambda^{(R)}$
for an interaction with maximum range $R$  can be rewritten as the
sum of two contributions: a Lieb-Robinson bound $\lambda^{(R')}$
for an interaction of maximum range $R'$, which can be interpreted as the “short-range part of the evolution”;
and an additional propagation of quantum information due to  “long-range hops”, which have range between $R'$ and $R$ and maximum strength $f(R')$. 
However, these ``long-range hops'' need not originate in the support of the original $A$
itself but, rather, in the support of the time-evolved $A$
under the short-range part of the interaction.
The latter spreads the quantum information by
${\mathcal I}[\lambda^{(R’)}]$, as depicted in Fig. \ref{fig:iteration}.

At each iteration we replace the short-range contribution by the short-range Lieb-Robinson bound.
We make use of the bound proven in Theorem A.1 of Ref.~\cite{Matsuta_1604} which state that, for observables $A$ and $B$ supported on sets $X$ and $Y$, respectively:
\begin{equation}\label{eq:short_Lieb}
\| [ \tau_t^{H^{\le R'}}(A), B] \| \leq 2  |X| \exp[vt -  d(X,Y)/R'] \| A \| \| B \|~.
\end{equation}

Finally, we are free to choose $R'$ in \eqnref{eqn:lemma-recursion} so we take it to be a function of $r$; specifically, at the $n$-th iteration we take $R' = r^{\sigma_n}$, with $d/\alpha < \sigma_n < 1$. The resulting bound no longer depends on any cut-off $R'$ and when used again in \eqnref{eqn:lemma-recursion}, leads to a faster decaying $\mathcal{I}[\lambda]$ and an improved bound.

Therefore, at the $n$-th iteration we obtain the bound:
\begin{equation}
\| [\tau_s^{H^{\le R}}(A), B] \| \leq \lambda^{(R)}_n(d(X,Y)) \| A \| \| B \|,
\end{equation}
where the iteration equation is:
\begin{equation}\label{iteration}
\lambda^{(R)}_n(r) = \Delta_{r\,} \left( 2|X| \exp\left[ vt - r^{1-\sigma_n}\right] + C \Theta(R - r^{\sigma_n}) |X| t\, f(r^{\sigma_n}) \mathcal{I}[\lambda_{n-1}^{(r^{\sigma_n})}] \right),
\end{equation}
where
\begin{equation}
  \Delta_{r\,}\!(u) = \begin{cases}
    2 & r < 1 ~\mathrm{or}~ u > 2 \\ 
    u & \mathrm{otherwise}
  \end{cases}
\end{equation}
This choice of $\Delta_r$ ensures that we always use the trivial bound on the commutator when $r=0$ or when it is the most stringent bound.
Now, it only remains to carry out the iteration.

\subsection{Analyzing the iteration}

To begin the iterative process we can invoke the
generic Lieb-Robinson bound for finite-range Hamiltonians, as described in \eqnref{eq:short_Lieb}.
Taking into account the trivial case,
\begin{equation}
\| [ \tau_t^{H^{\le R}}(A), B] \| \leq 2 \| A \| \| B \|,
\end{equation}
we begin the iteration with the initial bound:
\begin{equation}\label{eq:lambda0}
  \lambda_0^{(R)}(r) = \Delta_r(2|X| e^{vt-r/R}).
\end{equation}


\noindent
We then find (calculation in Appendix~\ref{app:I_n0}):
\begin{align}
\mathcal{I}[\lambda_0^{(R')}] \leq C|X|  \Bigl[ 1 + (vtR')^{d} \Bigr]  
\end{align}
%
Taking \eqnref{iteration} and setting $R'=r^{\sigma_1}$, we have:
\begin{equation}
  \label{lambda1}
  \lambda_{1}^{(R)}(r) \leq \Delta_{r\,\,} \!\!\Bigl( 2|X|e^{vt - r^{1-\sigma_1}} + C \Theta\bigl(R - r^{\sigma_1}\bigr) \,|X|^2 Jt r^{-\sigma_1 \alpha} [1 +
r^{\sigma_1d} (vt)^{d}]\Bigr)
\end{equation}
which recovers the results in Ref.~\cite{Matsuta_1604} with an appropriate choice of $\sigma_1$.
From this point, we proceed by induction. 
Indeed, suppose at the $n$-th iteration we have:
\begin{equation}
  \label{lambdan}
\lambda_n^{(R)}(r) \leq \Delta_{r\,\,} \!\!\Bigl(  2|X|e^{vt - r^{1-\sigma_n}}  + \,C \Theta(R - r^{\sigma_n}) \sum_{i=1}^{2} \mathfrak{F}^{(n)}_{i}(vt)\,
r^{\mu^{(n)}_i} \Bigr).
\end{equation}
Note that, according to \eqnref{lambda1}, this is satisfied for $n = 1$
if we take
\begin{align}
\mu^{(1)}_1 &=  \sigma_1(-\alpha + d) \label{eq:mu1_initial}\\
\mu^{(1)}_2 &=  -\sigma_1 \alpha\\
\mathfrak{F}^{(1)}_1(\tau) &= C \tau^{d+1} |X|^2\\
\mathfrak{F}^{(1)}_2(\tau) &= C \tau |X|^2 
\end{align}
Then, so long as $\mu_1^{(n)} + d > 0$ and $\mu_2^{(n)} + d < 0$
we have (as computed in Appendix \ref{app:I_n}):

%

\begin{equation}\label{eq:mathcalI_1}
\mathcal{I}[\lambda_n^{(R')}] \leq C \left[ \left\{|X|( 1 + (vt)^{d/(1-\sigma_n)}) + \mathfrak{F}_2^{(n)}(vt)~(vt)^{(d+\mu_2^{(n)})/(1-\sigma_n)}\right\} + \mathfrak{F}^{(n)}_{1}(vt)~ (R')^{(\mu^{(n)}_1 + d)/\sigma_n}\right]
\end{equation}
and therefore, using \eqnref{iteration} and setting $R' = r^{\sigma_{n+1}}$:
\begin{align} \label{eq:lambda_n}
  &\lambda_{n+1}^{(R)}(r) \leq \Delta_{r\,}\!\Bigl( 2|X|e^{vt - r^{1-\sigma_{n+1}}} +~\notag\\
  &+C \Theta(R - r^{\sigma_{n+1}}) |X| J t  r^{-\sigma_{n+1} (\alpha - (\mu^{(n)}_1 + d) /\sigma_n)}~  \mathfrak{F}^{(n)}_{1}(vt) + \notag \\
  &+C \Theta(R - r^{\sigma_{n+1}}) |X| J t  r^{-\sigma_{n+1} \alpha} \left\{|X|( 1 + (vt)^{d/(1-\sigma_{n})}) + \mathfrak{F}_2^{(n)}(vt)~(vt)^{(d+\mu_2^{(n)})/(1-\sigma_n)}\right\} ~.
\end{align}
By choosing $\sigma_{n+1} \leq \sigma_n$ we ensure that the spatial decay of the exponential term does not increase in performing the iterative procedure.

So at the next iteration we have
\begin{align}
  \label{starting_firsteq}
\mu^{(n+1)}_1 &= \sigma_{n+1}\bigl(-\alpha + (\mu^{(n)}_1 + d)/\sigma_n \bigr)\\
\mu^{(n+1)}_{2} &= -\sigma_{n+1} \alpha \\
\mathfrak{F}_1^{(n+1)}(\tau) &= C \tau|X| ~\mathfrak{F}_1^{(n)}(\tau)  \\
\mathfrak{F}_{2}^{(n+1)}(\tau) &= C \tau|X|\left\{ |X|[ 1 + \tau^{d/(1-\sigma_n)} ] + \mathfrak{F}_2^{(n)}(\tau)~ \tau^{(d+\mu_2^{(n)})/(1-\sigma_n)}\right\} 
  \label{starting_lasteq}
\end{align}

Iteratively applying \eqnref{starting_firsteq} to the initial condition of \eqnref{eq:mu1_initial} yields:
\begin{equation}
\label{eq:amu}
\mu^{(n)}_1 = \left(1+ \frac{1}{\sigma_1} +
\frac{1}{\sigma_2}+ \ldots \frac{1}{\sigma_{n-1}}\right){\sigma_n}d\, - \,
n{\sigma_n}\alpha.
\end{equation}

At each iteration, $\mu_1^{(n)}$ is made smaller (i.e. more negative)
at the cost of increasing the leading 
power of $\tau$ in $\mathfrak{F}_{1}^{(n)}(\tau)$, so long as $\mu^{(n)}_1 > -d$. 
By choosing appropriate $\sigma_j$, we eventually reach an iteration step $n=n^*$ such
that $\mu_1^{(n^*)} + d < 0$ and \eqnref{eq:mathcalI_1} no longer holds
(and neither will the iteration equations Eqs.~(\ref{starting_firsteq}) - (\ref{starting_lasteq})). 
For $n > n^*$, $\mathcal{I}[\lambda^{R'}_n]$ becomes independent of $R'$:
\begin{align}\label{eq:mathcalI_2}
  \mathcal{I}[\lambda_{n\ge n^*}^{R'}] \le C \left[ |X|(1 + (vt)^{d/(1-\sigma_n)}) + \mathfrak{F}_1^{(n)}(vt)~(vt)^{(d+\mu^{(n)}_1)/(1-\sigma_n)} + \mathfrak{F}_2^{(n)}(vt)~(vt)^{(d+\mu^{(n)}_2)/(1-\sigma_n)}\right]
\end{align}
which leads to new iterative steps where the spatial decay of both polynomial terms is the same: 
\begin{align}
  \mu^{(n+1)}_1 &= \mu^{(n+1)}_{2} = -\sigma_{n+1}\alpha\\
  \mathfrak{F}_1^{(n+1)}(\tau) &= \tau^{1+(d+\mu_1^{(n)})/(1-\sigma_n)}|X| ~\mathfrak{F}_1^{(n)}(\tau)  \label{eq:it_F1_new}\\
  \mathfrak{F}_{2}^{(n+1)}(\tau) &= \tau|X|\left\{ |X|[ 1 + \tau^{d/(1-\sigma_n)} ] + \mathfrak{F}_2^{(n)}(\tau) \tau^{(d+\mu_2^{(n)})/(1-\sigma_n)}\right\} ~.\label{eq:it_F2_new}
\end{align}

At this point in the iterative procedure,
further iterations do not improve on the power-law decay of the Lieb-Robinson bound since they are set by $-\sigma_{n} \alpha$.

With regards to the time dependence of the bound, at each iteration step $n$, one can choose $\sigma_n > (1-\sigma_{n-1} +d)/\alpha$, reducing the time dependence of $\mathfrak{F}_i^{(n)}(vt)$ in Eqs.~(\ref{starting_lasteq}), (\ref{eq:it_F1_new}) and (\ref{eq:it_F2_new}).
For such choices of $\sigma_n$ and enough iterations steps, the leading temporal dependence arises from the $\tau^{1+d/(1-\sigma_n)}$ term introduced each iteration step in \eqnref{eq:it_F2_new}.
As a result, there is some iteration number $m>n^*$ above which the most meaningful terms of the bound do not change.
At this point, the bound $\lambda^{(R)}_m(r)$ is given by:
\begin{equation}
  \lambda^{(R)}_m \le \Delta_r\left(2|X| e^{vt - r^{1-\sigma_1}} + C \Theta(R-r^{\sigma_m}) r^{-\sigma_m \alpha}\left\{ |X|^{2}(vt)^{1 + d/(1-\sigma_1)} + \ldots\right\} \right)~,
\end{equation}
where $\ldots$ are terms with lower power in $vt$, but higher power in $|X|$.

We can make the previous considerations more concrete by analyzing the case where $\sigma_j$ are all made equal, $\sigma_j = \sigma > (d+1)/(\alpha+1)$.
This inequality ensures the reduction of the time dependence of $\mathfrak{F}^{(n)}_i(vt)$.

For this choice of $\{\sigma_j\}$, \eqnref{eq:amu} simplifies to:
\begin{align}
  \mu^{(n)}_1 = (n-1+\sigma)d - n\sigma \alpha
\end{align}
further leading to $n^* = \lceil\sigma d/(\sigma\alpha-d)\rceil$.

For $n>n^*$, the time dependence is encoded in:
\begin{align}
  \mathfrak{F}_1^{(n)}(\tau) &\sim \tau^{1 + d/(1-\sigma)}  \left[\tau^{[1+d-\sigma(1+\alpha)]/[1-\sigma]}\right]^{n-1} + \ldots\\
  \mathfrak{F}_2^{(n)}(\tau) &\sim \tau^{1+ d/(1-\sigma)} + \ldots 
\end{align}
where $\ldots$ correspond to lower power of $\tau$. Then, $\mathfrak{F}_2(\tau)$ becomes the dominant term immediately for iteration step $n^*+1$ as the term $[~\cdot~]^{n-1}$ reduces the leading term of $\mathfrak{F}_1^{(n)}(\tau)$ to be smaller than $\mathfrak{F}_2^{(n)}(\tau)$.
Because different terms have different dependences on $|X|$, to ensure all constants are independent of $|X|$, we include the largest power of $|X|$ emmerging from our construction in front of the time dependence.
Finally, taking $R\to\infty$ yields the final result as expressed in Theorem \ref{thm:thethm}.

\section{Power-law light-cones}

In short-range interacting systems, the length scale associated with the exponential decay of the Lieb-Robinson bound, \eqnref{eq:short_Lieb}, provides a natural definition for a light-cone.
In contrast, Lieb-Robinson bounds in long-range interacting systems are characterized by power-law decays that lack a natural length scale\footnote{Such power-lay decays are present in current Lieb-Robinson bounds for long-range interacting systems, both in Refs~\cite{Hastings_0507,FossFeig_1410,Matsuta_1604} and this work.}.
As a result, the precise notion of a light-cone will vary depending on which properties we wish to capture.

One way to define a light-cone is in terms of the ``spread of information'': that is, suppose we consider the time evolution of two states $\ket{\psi}$ and $O \ket{\psi}$, where $O$ is a local operator that perturbs the initial state.
The light-cone is the region of radius $R_{\mathrm{LC1}}(t)$ around the support of $O$, outside which, both time-evolved states yield nearly identical local observables.
It is a direct measure of the spread of the influence of the perturbation $O$ across the system as a function of time $t$.
We refer to this light-cone as LC1.

A different way to define a light-cone is in terms of the region of the system that can affect the evolution of local observables appreciably.
More specifically, consider the time evolution of an operator $O$ under two different Hamiltonians, $H$ and $H+\Delta H$.
Intuitively, if $\Delta H$ only acts very far away from $O$, it will not have a significant impact on the evolution of $O$ at short times.
One can make this intuition precise and guarantee that the evolution of $O$ does not change appreciably, until time $t$, if $\Delta H$ only acts a distance $R_{\mathrm{LC2}}(t)$ away from $O$.
$R_{\mathrm{LC2}}(t)$ then characterizes the ``zone of influence'' of the evolution of operator $O$.
We refer to this light-cone as LC2.
Strictly speaking, LC2 is not a light-cone.
However, this ``zone of influence'' is intimately connected with a \emph{modified} notion of the past light-cone.
Our usual understanding of such a past light-cone consists of all events (points in space-time) where acting with a \emph{local operator} can influence the current event.
The modified past light-cone that is naturally associated with LC2 corresponds to all events where a change in the \emph{Hamiltonian} can influence the current event.
In long-range systems, these two light-cones need not be equal, as even a local change to the Hamiltonian can affect the system non-locally.

In general, in power-law interacting systems, LC2 will be greater than LC1.
Intuitively, as the operator $O$ expands outwards, the number of terms of $\Delta H$ it can interact with increases dramatically.
As a result, it is not only necessary that the operator is mostly localized to a particular region, but also that the spatial profile of the operator spread decays fast enough to counteract the increasing number of terms that can modify its dynamics.

We now make these definitions more precise.
In order to simplify the notation in this section, we write the Lieb-Robinson bound between two operators $A$ and $B$,
such that $d(A,B) = r$, as:
\begin{equation}
\label{eqn:LR-bound-general-form}
  \frac{\|[\tau^H_t(A),B]\|}{\|A\|\|B\|} \le \mathcal{C}(r,t),
\end{equation}
This allows us to formally define LC1 as the light-cone used in the previous literature:
\begin{definition}
  Let {\bf light-cone 1 (LC1)} be defined as a relation $r = f(t)$ such that:
  \begin{equation}\label{eq:LC1}
    \lim_{t\to\infty} \mathcal{C}(f(t), t) = 0.
  \end{equation}
\end{definition}

The meaning of LC1 is that the propagation of quantum information outside the light-cone is small and gets smaller as $t\rightarrow \infty$
\footnote{Let us note that this has been measured in different forms in different results.
  More specifically, Ref.\cite{FossFeig_1410} requires the probe operator $B$ to be localized at one site, measuring the spread of quantum information to any one site, while in Ref.\cite{Matsuta_1604} the probe operator need not be local.
  Our bound follows this second convention, more common in previous literature.}.
Because we are interested in the asymptotic behavior, we focus on power-law light-cones, $f(t) = t^{\gamma}$, which
characterize the Lieb-Robinson bounds considered here.
The smallest light-cone is characterized by the exponent $\beta^{\mathrm{LC1}}$, the infimum of the $\gamma$ which satisfy \eqnref{eq:LC1}.

In contrast we wish to define LC2 as the region outside which changing the Hamiltonian of the system has no significant impact in the evolution of the operator.
To obtain a precise condition for LC2, we consider how changing the Hamiltonian $H$ to $H+\Delta H$ impacts the evolution of an operator.
More specifically, we consider modifying the Hamiltonian only a distance $r_{min}$ away from the operator of interest $O$.
In \refapp{app:boundDifference}, we show the difference in the time evolved operators is bounded by:
\begin{equation}\label{eq:BoundDiff}
  \|e^{iHt}Oe^{-iHt} - e^{i(H+\Delta H)t} O e^{-i(H+\Delta H)t}\| \le C\Delta J\|O\|  t\int_{r_{min}}^\infty dr~r^{d-1} \mathcal{C}(r,t)
\end{equation}
where $\Delta J$ quantifies the difference between the Hamiltonians.

LC2 is then given by the relationship between $r_{min}$ and $t$ that ensures that operator difference, bound in \eqnref{eq:BoundDiff}, remains small and goes to zero in the long time limit.
This immediately motivates the definition of LC2 as follows:
\begin{definition}
  Let {\bf light-cone 2 (LC2)} be defined as a relation $r = f(t)$ such that:
  \begin{equation}\label{eq:LC2}
    \lim_{t\to\infty}\quad t \int_{f(t)}^\infty dr~r^{d-1} \mathcal{C}(r, t) = 0,
  \end{equation}
  where $d$ is the dimensionality of the system.
\end{definition}

Again, we will focus on polynomial light-cones, $f(t) = t^\gamma$ and define $\beta^{\mathrm{LC2}}$ as the infimum of the $\gamma$ which satisfy \eqnref{eq:LC2}.

In short-ranged interacting systems, where
$\mathcal{C}(r, t) \propto e^{vt-r/R}$, the exponential suppression of $\mathcal{C}(r,t)$ at large $r$ is insensitive to the extra volume term in the definition of LC2, \eqnref{eq:LC2}, leading to the same linear light-cone  for both LC1 and LC2. 
This result is an immediate consequence of the natural length scale in $\mathcal{C}(r,t)$.

However, in long-range interacting systems, $\mathcal{C}(r,t)$ has a power-law decay which is sensitive to the extra volume term in LC2.
For example, for \eqnref{eq:LC2} to converge and ensure a power-law LC2, the Lieb-Robinson bound must decay \emph{faster} than $r^{-d}$; for LC1 there is no such requirement.
As a result, for slowly-decaying Lieb-Robinson bounds one may have a power-law LC1 but no LC2, i.e. there is no power-law $f(t)$ that satisfies \eqnref{eq:LC2}.
This is the case for
the bound in Matsuta et al. \cite{Matsuta_1604}, where LC2 does not exist for $d<\alpha<2d$, yet LC1 matches that of Foss-Feig et al. \cite{FossFeig_1410}.
LC2 is able to capture the difference between these two results.

By comparison, our result supports both an LC1 and LC2 for $d < \alpha$, extending the existence of an LC2 in long-range multi-spin Hamiltonians to $d<\alpha<2d$.
In this regime both our Lieb-Robinson bound and that of Ref.~\cite{FossFeig_1410}
lead to finite LC2, albeit with a larger light-cone exponent for our bound.
Much like the difference in decay profile, this might be inherent to our treatment of the more general case of arbitrary multi-spin interactions.

\begin{figure}

  \centering
  \includegraphics[width=0.7\linewidth]{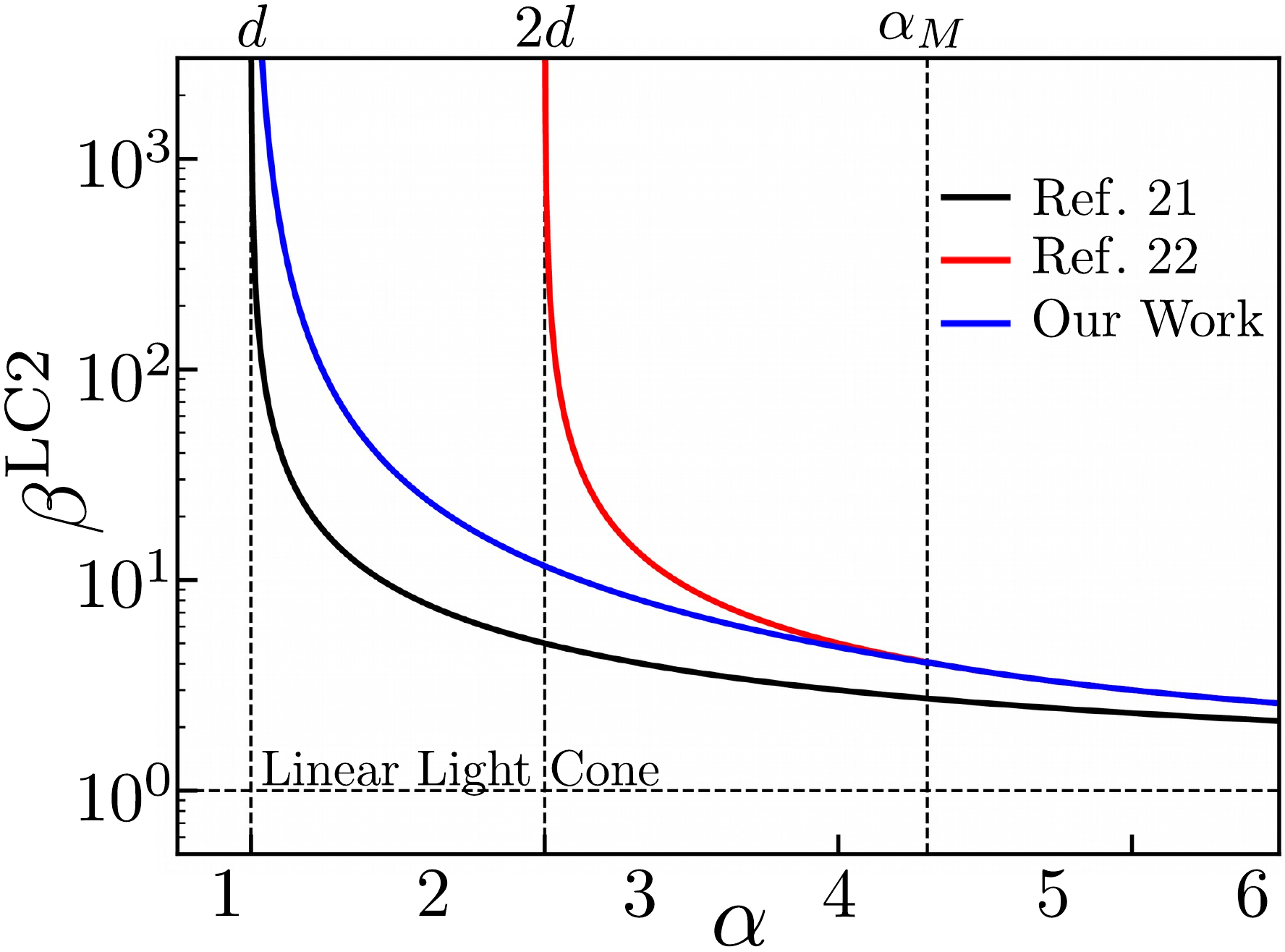}
  \caption{
    Power-law LC2 exponent for the present paper and Refs. \cite{FossFeig_1410} and \cite{Matsuta_1604} for $d=1$ as a function of $\alpha$.
    While Ref.~\cite{Matsuta_1604} has a finite power-law LC2 for $\alpha > 2d$, Ref.~\cite{FossFeig_1410} and our work have a power-law LC2 for all $\alpha > d$.
    For $\alpha < \alpha_M$ our work leads to a better LC2 than Ref.~\cite{Matsuta_1604}, while matching it for $\alpha \ge \alpha_M$.
    The horizontal dashed line corresponds to a linear light-cone.
    More details about the calculation can be found in Appendix \ref{sec:appendix_light_cone}.
  }
  \label{fig:LC2_d1}
\end{figure}

In Table \ref{light-cone-table} and Figure \ref{fig:LC2_d1}, we compare the different light-cone exponents obtained from both our work and previous literature for different values of $\alpha$.
In Figure \ref{fig:LC2_d1}, we plot the exponent of LC2 of the different works as a function of $\alpha$ for dimension $d=1$.
More details on the calculation, and the general formulae for all space dimensions $d$, can be found in Appendix \ref{sec:appendix_light_cone}.

{\renewcommand{\arraystretch}{2.3} 
  \begin{table}
    \begin{adjustwidth}{-.5in}{-.5in}  
      \begin{center}
        \begin{tabular}{|c|c|c|c|c|} \hline
          Reference & LC1 $(d<\alpha)$ &LC2 $(d < \alpha \le 2d)$ & LC2 $\left(2d< \alpha < \alpha_M\right)$& LC2 $\left( \alpha_M \le \alpha \right)$ \\ \hline \hline
          Ref.~\cite{FossFeig_1410} & $\beta^{\mathrm{LC1}}_{\text{FF}} = \dfrac{\alpha+1}{\alpha-d}$ &$\beta^{\mathrm{\mathrm{LC2}}}_{\text{FF}} = \dfrac{\alpha + d}{\alpha}\dfrac{\alpha+1}{\alpha-d}+\dfrac{1}{\alpha}$ & $\beta^{\mathrm{LC2}}_{\text{FF}} = \dfrac{\alpha + d}{\alpha}\dfrac{\alpha+1}{\alpha-d}+ \dfrac{1}{\alpha}$ & $\beta^{\mathrm{LC2}}_{\text{FF}} = \dfrac{\alpha + d}{\alpha}\dfrac{\alpha+1}{\alpha-d}+\dfrac{1}{\alpha}$\\\hline
          Ref.~\cite{Matsuta_1604} &$\beta^{\mathrm{LC1}}_{\text{M}} = \dfrac{\alpha+1}{\alpha-d}$ &\xmark  & $\beta^{\mathrm{LC2}}_\text{M} = \dfrac{\alpha + 2}{\alpha - 2d}$ & $\beta^{\mathrm{LC2}}_\text{M} = \dfrac{\alpha + 2}{\alpha - 2d}$ \\\hline
          Present Work &$\beta^{\mathrm{LC1}} = \dfrac{\alpha+1}{\alpha-d}$ & $ \beta_{\text{FF}}^{\mathrm{LC2}} < \beta^{\mathrm{LC2}} = \tilde{\beta} $ & $ \beta_{\text{FF}}^{\mathrm{LC2}} < \beta^{\mathrm{LC2}}= \tilde{\beta} <\beta^{\mathrm{LC2}}_{\text{M}}  $& $\beta^{\mathrm{LC2}} = \dfrac{\alpha+2}{\alpha-2d}$ \\\hline
        \end{tabular}
        \caption{Summary of the power-law light-cone exponents of LC1 and LC2 for both previous literature and our work. We use the subscript FF and M to refer to the light-cone exponents from the bounds of Refs.~\cite{FossFeig_1410} and \cite{Matsuta_1604} respectively.
          Here $\tilde{\beta} = \frac{2}{(\alpha-d)^2}\times \left[\alpha - d + \alpha d(1+ \sqrt{1 + 2/d - 2/\alpha})\right]$ and  $\alpha_M = \frac{3d}{2}\left[ 1 + \sqrt{ 1+\frac{8}{9d}} \right]$. For a detailed calculation see Appendix \ref{sec:appendix_light_cone}.}
        \label{light-cone-table}
      \end{center}
      \end{adjustwidth}
  \end{table}
}

\section{Discussion}
In this paper, we have proven an improved Lieb-Robinson bound for generic multi-body long-range interactions, characterized by a faster asymptotic spatial decay. 
The importance of this improvement is captured by the notion of LC2, a definition of light-cone that provides a stricter definition of locality for the growth of operators, in particular, that their evolution is not affected by the outside region for large $t$.
Our work extends the existence of an LC2 light-cone for generic multi-body interacting systems for $d<\alpha < 2d$.

This improvement has important implications for understanding prethermalization and Floquet phases of matter in periodically-driven systems. In such systems (especially in the high frequency regime), one can capture the evolution under a time-dependent Hamiltonian $H(t)$ using a time-independent approximation. Even when the original $H(t)$ has strictly two-body terms, the time-independent approximation will naturally exhibit multi-body terms.   
The results which establish the accuracy and limitations of such approximations require
Lieb-Robinson bounds for multi-body power-law interactions with a rapid decay
outside the light-cone.
This application will be discussed in future work \cite{Machado_18xx}.

During the preparation of this manuscript, the authors became aware of a new Lieb-Robinson bound \cite{Tran_1808} that improves upon Ref.~\cite{FossFeig_1410}.
The bound in Ref.~\cite{Tran_1808} has an LC1 exponent of $\alpha/(\alpha-d)$ under similar assumptions as Ref.~\cite{FossFeig_1410}, namely, two-body interactions.
However, their result (phrased in terms of commutators) does not yield a finite LC2 for $d < \alpha < 2d$.
Nevertheless, the structure of their arguments is intriguing and understanding how to generalize their results to multi-body interactions is a promising direction for future study.

\acknowledgements
We thank A. Gorshkov, G. Meyer, S. Michalakis, and C. Olund for helpful discussions. This work was supported by the NSF (PHY-1654740), the ARO STIR Program and the A.P. Sloan Foundation. D. Else was supported by the Microsoft Corporation and by the Gordon and Betty Moore Foundation.

\appendix
\section{Some technical results}
\begin{lemma}
  \label{lem:sumtointegral}
Let $f(r)$ be a monotonically decreasing function of $r$, and fix an $x \in \Lambda$. Then 
\begin{equation}
\sum_{z \in \Lambda, a \leq d(z,x) \leq R} f(d(z,x)) \leq \frac{C}{a^d} \int_{a/2}^{R} f(r) r^{d-1} ~dr,
\end{equation}
where $a$ is the minimum separation between lattice points.
\begin{proof}
Around each lattice point $z$, consider a ball $B_z$ of radius $a/2$. Given our
  assumption that $a$ is the smallest separation between lattice points, these balls are pairwise disjoint (up to sets of measure zero). Now, for any $\textbf{r}$ in $B_z$, we have that $f(|\textbf{r} - \textbf{r}_x| - a/2) \geq f(d(z,x))$. Therefore, 
\begin{equation}
V f(d(z,x)) \leq \int_{B_z} f(|\textbf{r} - \textbf{r}_x| - a/2) ~d^d \textbf{r},
\end{equation}
where $V$ is the volume of the ball $B_z$. In the case that $d(x,z) < 3a/2$, we will use the tighter bound
\begin{equation}
V f(d(z,x)) \leq \int_{B_z : |\textbf{r} - \textbf{r}_x| < a} f(|\textbf{r} - \textbf{r}_x|)~ d^d \textbf{r} + 
\int_{B_z : |\textbf{r} - \textbf{r}_x| > a} f(|\textbf{r} - \textbf{r}_x| - a/2) ~d^d \textbf{r},
\end{equation}
Now using the fact that $\cup B_z \subseteq \mathbb{R}^d$, we find that
\begin{align}
\sum_{z \in \Lambda : a \leq d(z,x) \leq R} f(d(z,x))
&\leq \frac{C}{a^d} \left( \int_{a/2}^{a} r^{d-1} f(r) ~dr + \int_{a}^{R+a/2} r^{d-1} f(r - a/2) ~dr\right).
\end{align}
We can bound the second integral by
\begin{align}
\int_{a}^{R + a/2} r^{d-1} f(r - a/2)~dr
&= \int_{a/2}^{R} (u + a/2)^{d-1} f(u)~du \\
&\leq C' \int_{a/2}^{R} u^{d-1} f(u)~du.
\end{align}
This immediately proves the Lemma.

\end{proof}
\end{lemma}

\begin{lemma}
  \label{lem:incompletegamma}
  For any $\mu$ and $\nu$ and positive $\rho$ then the following inequality holds for a constant $C$  independent of $\rho$:
\begin{equation}
\label{incompletegammaeq}
\int_\rho^{\infty} e^{- x^{\nu}} x^{\mu}~dx \leq  C e^{-\rho^\nu} \left(1 + \rho^{\mu - \nu + 1}\right)
\end{equation}
\begin{proof}
  It sufficient to consider the case of $\nu = 1$, since we can reduce to this case by a change of variables.
  Let us first consider $\mu < 0$, then:
\begin{align}
\int_\rho^{\infty} e^{-x} x^{\mu} ~dx\leq \rho^{\mu}\int_\rho^{\infty} e^{-x}~dx &= \rho^{\mu} e^{-\rho} 
\end{align}

We are now left with the case $\mu > 0$. In that case, if $\rho \leq 1$ then we can certainly bound the left-hand side of \eqnref{incompletegammaeq} by:
\begin{equation}
  \int_\rho^{\infty} e^{-x} x^\mu~dx \le e^{1-\rho} \left[ \int_0^\infty e^{-x} x^\mu \right] = C_1 e^{-\rho}.
\end{equation}
On the other hand, for $\rho \geq 1$, we have
\begin{align}
\int_\rho^{\infty} e^{-x} x^{\mu} ~dx&= e^{-\rho} \rho^\mu \int_\rho^{\infty} e^{-(x-\rho)} (x/\rho)^{\mu}~dx \\
&= e^{-\rho} \rho^\mu \int_0^{\infty} e^{-u} (u/\rho + 1)^{\mu}~du \\
&\leq e^{-\rho} \rho^\mu \int_0^{\infty} (u+1)^{\mu} e^{-u} ~du\\
&= C_2 e^{-\rho} \rho^\mu \le C_2 e^{-\rho} \rho^{\mu + \epsilon},
\end{align}
for any $\epsilon > 0$. Adding both bounds with $C = \max(C_1,C_2)$ ensures it holds for all values of $\rho$. \qedhere
\end{proof}
\end{lemma}

\section{Calculation of $\mathcal{I}[\lambda]$}
In this Appendix we perform the calculation of short-range spread of quantum information $\mathcal{I}[\lambda^{(R')}_n]$, defined in \eqnref{eqn:recursion-integral}.
We divide this calculation into two cases, $n=0$ and $n>0$, where $\lambda^{(R')}_n$ takes different functional forms.

\subsection{$n=0$ case}
\label{app:I_n0}

In the initial bound, given in \eqnref{eq:lambda0} of the main text, one can define a ``light-cone'' inside which the trivial bound is best, described by:
\begin{align}
  |X|e^{vt - r/R'} = 1 \quad \Rightarrow\quad  r = R'(\ln |X| + vt).
\end{align}

As a result, one can bound $\mathcal{I}[\lambda^{(R')}_0]$ by the less stringent ``light-cone'' $r = R'vt$ as follows:
\begin{align}
  \mathcal{I}[\lambda^{(R')}_0] &= \lambda^{(R')}_0(0) + \int_{1/2}^\infty \rho^{d-1}  \lambda^{(R')}_0(\rho) ~d\rho\\
  &\le 2 + 2 \int_{1/2}^{R'vt} \rho^{d-1}  ~d\rho + 2|X| \int_{R'vt}^\infty \rho^{d-1} e^{vt - \rho/R'}~d\rho\\
  &\le 2\left[ 1 + \frac{1}{d}(R'vt)^d + |X|e^{vt} e^{-vt}C (1+(R'vt)^{d-1})\right]\\
  &\le C\left[ |X| + |X|(R'vt)^{d-1} + (R'vt)^d \right]
\end{align}
where we made use of Lemma~\ref{lem:incompletegamma} to bound the second integral.

This bound can be made less stringent as follows:
\begin{align}
  \mathcal{I}[\lambda^{(R')}_0] &\le C|X|\left[ 1 + (R'vt)^{d}\right].
\end{align}

This simplification leads to one less polynomial term in our iterative analysis but does not affect the spatial or temporal asymptotic behavior of the bound.
In contrast, it increases $\mathfrak{F}_1$ by a factor of $|X|$ in our construction.

\subsection{$n\ge 1$ case}
\label{app:I_n}

For $n\ge 1$, the bound $\lambda^{(R')}_n(r)$ is composed of an exponential term and two polynomial terms, as described in \eqnref{lambdan} of the main text.

Similar to the calculation for $n=0$, there exists a ``light-cone'' inside of which the trivial bound is best.
Such ``light-cone'', in principle, will depend on the polynomial terms of the bound, however, it must be at least as big as the length scale of the exponential term of the bound given by:
\begin{equation}
  vt - r^{1-\sigma_n} = 0 \Rightarrow r = (vt)^{1/(1-\sigma_n)}
\end{equation}

One can then bound the spread of quantum information as:
\begin{align}
  &\mathcal{I}[\lambda^{(R')}_0] = \lambda^{(R')}_0(0) + \int_{1/2}^\infty \rho^{d-1}  \lambda^{(R')}_0(\rho) ~d\rho\\
  &\le 2 + 2 \int_{1/2}^{(vt)^{1/(1-\sigma_n)}} \rho^{d-1}  ~d\rho +\\
  &\hspace{1cm}+\int_{(vt)^{1/(1-\sigma_n)}}^\infty \rho^{d-1} \left[2|X|e^{vt - \rho^{1-\sigma_n}} + C \Theta(R'-\rho^{\sigma_n})\sum_{i=1}^2 \mathfrak{F}_i^{(n)}(vt) \rho^{\mu_i^{(n)}} \right]~d\rho\\
  &\le 2 + 2 \int_{1/2}^{(vt)^{1/(1-\sigma_n)}} \rho^{d-1}  ~d\rho + 2|X| \int_{(vt)^{1/(1-\sigma_n)}}^\infty \rho^{d-1} e^{vt - \rho^{1-\sigma_n}} ~d\rho
\end{align}
\begin{align}
  &\hspace{1cm}+\sum_{i=1}^2  C \int_{(vt)^{1/(1-\sigma_n)}}^{R'^{1/\sigma_n}} \rho^{d-1} \mathfrak{F}_i^{(n)}(vt) \rho^{\mu_i^{(n)}} ~d\rho\\
  &\le 2 + \frac{2}{d} (vt)^{d/(1-\sigma_n)} +  2|X| C[1 + (vt)^{ (d)/(1-\sigma_n)- 1}] + C \sum_{i=1}^2\mathfrak{F}_i^{(n)}(vt)  \left.\frac{\rho^{d + \mu_i^{(n)}}}{d+\mu_i^{(n)}}\right|_{(vt)^{1/(1-\sigma_n)}}^{(R')^{1/\sigma_n}}
\end{align}
The sign of $d+ \mu_i^{(n)}$ becomes important in bounding the polynomial terms
\footnote{In this analysis, we disregard the case where $d+\mu_i^{(n)}=0$ which would lead to logarightmic corrections. In the iterative construction of the main text, this can always be avoided by an appropriate choice of $\sigma_j$.} :
if $d+\mu_i^{(n)} > 0$, we can bound the term solely by the upper limit of integration, if $d + \mu_i^{(n)} < 0$, then we can bound using the lower limit.
The final bound on $\mathcal{I}[\lambda^{(R')}_n]$ then becomes:
\begin{align}
  \mathcal{I}[\lambda^{(R')}_n] &\le C(|X|+ |X|(vt)^{d/(1-\sigma_n)-1} + (vt)^{d/(1-\sigma_n)})  \\
  &+ C \sum_{i=1}^2\mathfrak{F}_i^{(n)}(vt) \times 
  \begin{cases}
    (R')^{(d+\mu_i^{(n)})/\sigma_n} & d+\mu_i^{(n)} > 0\\
    (vt)^{(d+\mu_i^{(n)})/(1-\sigma_n)} & d+\mu_i^{(n)} < 0
  \end{cases}
\end{align}

This bound can be slightly simplified at the expense of a higher dependence on $|X|$ on the $(vt)^{d/(1-\sigma_n)}$ term.
Nevertheless, this simplification does not change the asymptotic spatial or temporal decay of our results:
\begin{align}
\mathcal{I}[\lambda^{(R')}_n]  &\le C|X|( 1+ (vt)^{d/(1-\sigma_n)}) + C \sum_{i=1}^2\mathfrak{F}_i^{(n)}(vt) \times 
  \begin{cases}
    (R')^{(d+\mu_i^{(n)})/\sigma_n} & d+\mu_i^{(n)} > 0\\
    (vt)^{(d+\mu_i^{(n)})/(1-\sigma_n)} & d+\mu_i^{(n)} < 0
  \end{cases}
\end{align}


\section{Bounding the operator difference under two different Hamiltonians}
\label{app:boundDifference}

Consider a local operator $O$, which is time evolved under two different Hamiltonians $H_1$ and $H_2$.
Let us consider $\Delta H = H_1 - H_2$ such that it is only non-zero at sites outside some radius $r_{min}$ around $O$ and quantify its difference in terms of the largest local difference:
\begin{align}
  \sup_{x \in \Lambda} \sum_{Z: x\in Z} \|(H_1)_Z - (H_2)_Z\| = \sup_{x \in \Lambda} \sum_{Z: x\in Z} \| \Delta H_Z\| < \Delta J,
\end{align}
where $\Lambda$ is the set of sites of the system and $\| \cdot \|$ corresponds to the norm of the $H_Z$ term.

The goal of this section is to bound how much $O$ will differ when evolved under the two different Hamiltonians.
More particular, we consider the following norm:
\begin{equation}
  \| U_1^\dag O U_1 -  U_2^\dag O U_2\| = \| O - U_1 U_2^\dag O U_2 U_1^\dag\| \quad \text{ where } \quad U_n = e^{-iH_n t}
\end{equation}
where the time dependence of $U_n$ is implicit to simplify the notation.
Let us also note the similarities to results in Loschmidt echoes, where one evolves the system forwards with one Hamiltonian $H_1$ and then backwards with a slightly different Hamiltonian $H_2$ \cite{Peres__8410,Jalabert_0002}.

We begin by noting the following property:
\begin{align}
  f(t) &= O - U_1(t) U_2^\dag(t) O U_2(t) U_1^\dag(t)\\
  &\quad \Rightarrow \quad \frac{d}{dt} f(t) = -iU_1(t) [U_2(t)OU_2^\dag(t),\Delta H] U_1^\dag(t)
\end{align}
where we used the fact that $[U_n,H_n] = 0$.
One can now bound the difference as:
\begin{align}
  \|f(t)\| &= \left\| \int_0^t ds ~iU_1(s) [U_2(s)OU_2^\dag(s), \Delta H] U_1^\dag(s)\right \| \le \int_0^t ds ~\left \| [U_2(s)OU_2^\dag(s), \Delta H]\right \|\\
  &\le \int_0^t ds ~\sum_Z\left \| [U_2(s)OU_2^\dag(s), \Delta H_Z]\right \| 
\end{align}

We now focus our attention to the inner sum. Because $\Delta H$ is only non-zero on sites at $r_{min}$ away from the operator $O$ we can bound:
\begin{align}
  \sum_Z&\left\| [U_2(s)OU_2^\dag(s), \Delta H_Z]\right\| \le \sum_{z:d(z,O) \ge r_{min}}  \sum_{\substack{Z: z\in Z, \\ d(Z,O) =d(z,O)}}\left\| [U_2(s)OU_2^\dag(s), \Delta H_Z]\right\| \label{eq:C7}\\
  &\le \sum_{z:d(z,O) \ge r_{min}}\sum_{\substack{Z: z\in Z \\ d(Z,O) =d(z,O)}} \|O\|  \|\Delta H_Z\| \mathcal{C}(d(O,z),s) \label{eq:C8}\\
  &= \|O\|\sum_{z:d(z,O) \ge r_{min}} \mathcal{C}(d(O,z),s)  \sum_{\substack{Z: z\in Z \\ d(Z,O) =d(z,O)}} \|\Delta H_Z\| \\
  &\le \|O\|\sum_{z:d(z,O) \ge  r_{min}} \mathcal{C}(d(O,z),s) \Delta J \quad \le \quad  C \|O\| \Delta J \int_{r_{min}}^\infty dr~r^{d-1} \mathcal{C}(r,s)~,
\end{align}
where we used Lemma~\ref{lem:incompletegamma} to turn the sum into an integral.

Using the fact that $\mathcal{C}(r,s)$ is an increasing function in $s$ we can obtain the final bound:
\begin{align}
  \| U_1^\dag O U_1 -  U_2^\dag O U_2\| \le C\|O\| \Delta J t \int_{r_{min}}^\infty dr~r^{d-1}\mathcal{C}(r,t)
\end{align}

\section{Calculation of Light-cones}
\label{sec:appendix_light_cone}

Our task in this Appendix is to determine the LC1 and LC2 light-cones for Refs.~\cite{FossFeig_1410},\cite{Matsuta_1604} and our work.
In order to simplify the notation, let us write the Lieb-Robinson bounds in terms of $\mathcal{C}(r,t)$ as defined in \eqnref{eqn:LR-bound-general-form}.
For the different results, $\mathcal{C}(r,t)$ contains a combination of exponential
and power-law terms which need to be considered in determining LC1 and LC2.

Let us note that the iterative construction that leads to the bound in Theorem~\ref{thm:thethm} depends on two parameters: $\sigma$, the scaling of the inner cutoff in the iterative procedure, and $n$, the number of iterations performed.
While the fastest spatial decay occurs for $\sigma \rightarrow 1$, this does not
necessarily lead to the smallest light-cone, as the spatial decrease occurs at the expense of an increased growth in the temporal dependency.
The same is true for the number of iterations $n$.
As a result, one has to optimize both $\sigma$ and $n$ to find the smallest light-cone.

\subsection{Light-Cone 1 (LC1) for power-law interactions}

\subsubsection{Foss-Feig et al. Ref.~\cite{FossFeig_1410} and Matsuta et al. Ref.~\cite{Matsuta_1604}}

The computation of LC1 for Refs.~\cite{FossFeig_1410} and \cite{Matsuta_1604} is performed in their work, leading to a matching light-cone power-law:
\begin{align}
  \beta^{\mathrm{LC1}}_{\mathrm{FF}} = \beta^{\mathrm{LC1}}_{\mathrm{M}} = \frac{\alpha+1}{\alpha-d}~,
\end{align}
where the subscript FF and M refer to Refs.~\cite{FossFeig_1410} and \cite{Matsuta_1604} respectively.

\subsubsection{Our work}

As described in the main text, our proposed iterative construction matches the result of Ref.~\cite{Matsuta_1604} for $n=1$.
As a result, for $n=1$ we have $\beta^{\text{LC1}} = \beta^{\text{LC1}}_{\text{M}}$.

We now can show that performing further iterative steps does not change the value of the LC1 exponent.
In the iterative construction of our Lieb-Robinson bound, for $n>1$ and $\sigma_j = \sigma$ for all $j$, we have:
\begin{align}
  \mathcal{C}(r,t) \le C\left\{e^{vt - r^{1-\sigma}} + \mathfrak{F}_1^{(n)}(vt) r^{\mu_1^{(n)}} + \mathfrak{F}_2^{(n)}(vt) r^{\mu_2^{(n)}}\right\}~.
\end{align}
where we have absorbed any $|X|$ dependence into the constant $C$ as it does not affect the light-cone calculation; and:
\begin{align}
  \mathfrak{F}_i^{(n)}(\tau) = \tau^{\gamma_i^{(n)}} + \ldots
\end{align}
where $\ldots$ refers to lower power of $\tau$. Because we are interested in the late time asymptotic form of the light-cone we only need to focus on the largest power of $\tau$.
This exponent $\gamma_i^{(n)}$ is given by:
\begin{align}
  \gamma_1^{(n)} &= 
  \begin{cases}
    d+n & n \le n^*\\
    \dfrac{\sigma \alpha + n(1+d-\sigma(1+\alpha))}{1-\sigma} & n > n^* 
  \end{cases} \label{eq:gamma1}\\
  \gamma_2^{(n)} &= 1+\frac{d}{1-\sigma} + \max\left[0 ~,~ \frac{n-2}{1-\sigma}\left\{ 1 +d -\sigma(1+\alpha)\right\} \right]\label{eq:gamma2}
\end{align}
An important remark is that if $\sigma > (d+1)/(\alpha+1)$ then $1 + \frac{d-\sigma\alpha}{1-\sigma} < 0$.
In this regime, increasing $n$ reduces $\gamma_1^{(n)}$ for $n>n^*$ and does not change $\gamma_2$.

The spatial decay is then given by:
\begin{align}
  \mu_1^{(n)} &= \begin{cases}
    -n\sigma \alpha + d(\sigma + n-1) & n\le n^*\\
    -\sigma \alpha & n > n^*
    \end{cases}\label{eq:mu1}\\
  \mu_2^{(n)} &= -\sigma\alpha \label{eq:mu2}
\end{align}

Each of the three terms (the exponential and the two polynomials) will lead to a LC1 exponent.
The final exponent is the largest of the three for some $n$ and $\sigma$.
Optimizing over these two parameters yields the best $\beta^{\mathrm{LC1}}$.
\begin{align}
  \beta^{\text{LC1}}_{n;exp} = \frac{1}{1-\sigma} \quad,\quad \beta^{\text{LC1}}_{n;poly1} = \frac{\gamma_1^{(n)}}{-\mu_1^{(n)}}\quad,\quad \beta^{\text{LC1}}_{n;poly2} = \frac{\gamma_2^{(n)}}{-\mu_2^{(n)}}
\end{align}

Immediately, one can see that $\beta^{\text{LC1}}_{n;exp}$ is an increasing function of $\sigma$.
At the same time $\beta^{\text{LC1}}_{n;poly1}$ is a decreasing function of $\sigma$ for any fixed $n$, as shown below.
The intersection of the two curves provides the best LC1 exponent from the exponential and first polynomial term alone.
This intersection occurs at $\sigma = \sigma_{1;exp}$.
If $\beta^{\text{LC1}}_{n;poly2}(\sigma_{1;exp})$ is less or equal than the other two curves at this point, it corresponds to the correct LC1 exponent.

We begin by showing that $\beta^{\text{LC1}}_{n;poly1}$ is a decreasing function. If $n\le n^*$:
\begin{align}
  \beta^{\text{LC1}}_{n;poly1} = \frac{d+n}{ \sigma(n \alpha - d) - (n-1)d}
\end{align}
which is a decreasing function of $\sigma$.

We now focus on the case $n>n^*$:
\begin{align}
  \beta^{\text{LC1}}_{n;poly1} = \frac{\sigma \alpha + n(1+d -\sigma(1+\alpha))}{ (1-\sigma)\sigma\alpha}
\end{align}
First, let us note that $n^*$ is a decreasing function of $\sigma$. Moreover, for this calculation to be meaningful we need:
\begin{align}
  n>  n^*(\sigma=1)\Rightarrow n \ge n^*(\sigma=1) + 1 =  \left\lceil \frac{ d}{\alpha-d}\right\rceil + 1 = \left\lceil \frac{\alpha}{\alpha-d}\right\rceil \Rightarrow n \ge \frac{\alpha}{\alpha-d}
\end{align}
We can now compute the derivative of $\beta^{\text{LC1}}_{n;poly1}$ with respect to $\sigma$:
\begin{align}
  \frac{d}{d\sigma}\beta^{\text{LC1}}_{n;poly1} = \frac{\alpha\sigma^2 - n[(1+d)(1-2\sigma) + \sigma^2(1+\alpha)]}{\alpha \sigma^2 (1-\sigma)^2}
\end{align}
Parameterizing $\alpha = d+\epsilon$ and $n = \alpha/(\alpha-d) + \delta = (d+\epsilon)/\epsilon + \delta$, with $\epsilon>0 $ and $\delta \ge 0$, we obtain:
\begin{align}
  \frac{d}{d\sigma}\beta^{\text{LC1}}_{n;poly1} = -\frac{1}{\epsilon}~ \frac{1+d}{\sigma^2} - \delta \frac{ (1+d)(1-\sigma)^2 + \sigma^2 \epsilon}{(d+\epsilon)\sigma^2 (1-\sigma)^2}
\end{align}
which is always negative. Since the function is continuous, and in both cases it is decreasing, it is always decreasing.

The intersection of the two curves then occurs at:
\begin{align}
  \text{If }n\le n^* \quad&:\quad  \frac{d+n}{\sigma(n\alpha-d) - (n-1)d} = \frac{1}{1-\sigma} \Rightarrow \sigma_{1;exp} = \frac{d+1}{\alpha+1}\\
  \text{If }n> n^* \quad&:\quad  \frac{\sigma \alpha + n(1+d-\sigma(1+\alpha))}{\sigma \alpha (1-\sigma)} = \frac{1}{1-\sigma} \Rightarrow \sigma_{1;exp} = \frac{d+1}{\alpha+1}
\end{align}
which regardless of the regime occurs at the same value of $\sigma$, leading to:
\begin{align}
  \beta^{\text{LC1}}_{exp;poly1} = \frac{\alpha + 1}{\alpha-d}.
\end{align}

At the same time:
\begin{align}
  \beta^{\text{LC1}}_{n;poly2}(\sigma =\sigma_{1;exp} ) = \frac{\alpha+1}{\alpha-d} = \beta^{\text{LC1}}_{exp;poly1} = \beta^{\text{LC1}}
\end{align}
which corresponds to the best LC1 light-cone for this bound (equal for any number of iterations), in agreement with the previous works \cite{FossFeig_1410, Matsuta_1604}.

\subsection{Light-Cone 2 (LC2) for power-law interactions}


\subsubsection{Foss-Feig et al. Ref.~\cite{FossFeig_1410}}

We can summarize the bound obtained in Ref.~\cite{FossFeig_1410} as\footnote{As mentioned in footnote $^1$, we follow the notation of Ref.~\cite{Matsuta_1604} which differs from Ref.~\cite{FossFeig_1410} in its definition of $\alpha$. The ``$\alpha$'' in the latter should read $\alpha+d$ in the current notation.}:
\begin{align}
  \mathcal{C}_{\mathrm{FF}}(r,t) = \exp\left[ vt - \frac{r}{t^\gamma}\right] + \frac{t^{(\alpha+d)(1+\gamma)}}{r^{\alpha + d}}\quad \mathrm{where}\quad \gamma = \frac{1+d}{\alpha-d}
\end{align}

One can immediately extract the light-cone associated with LC2 for the two terms:
\begin{align}
  \beta^{\mathrm{LC2}}_{\mathrm{FF};exp} = \frac{\alpha+1}{\alpha-d} \quad,\quad \beta^{\mathrm{LC2}}_{\mathrm{FF};poly} = \frac{\alpha+d}{\alpha} \frac{\alpha+1}{\alpha-d} + \frac{1}{\alpha}
\end{align}
Since the latter is larger, it sets $\beta^{\mathrm{LC2}}_{\mathrm{FF}}$, which is valid for $\alpha > d$.

Let us note that because the Lieb-Robinson bound in Ref.~\cite{FossFeig_1410} holds only for two body interactions, the calculation of LC2 also only holds for such Hamiltonians $H_1$.
Moreover, because the bound is only valid for operators $A$ and $B$ which lie at a single site, the derivation in Appendix \ref{app:boundDifference} needs to consider the size of each term $H_z$.
More specifically, in going from line \eqnref{eq:C7} to \eqnref{eq:C8}, one should multiply each term by the size $Z$, the support of $H_Z$.
This does not affect the asymptotic behavior of the light-cone as long as:
\begin{align}
  \sup_{x\in \Lambda} \sum_{Z:x\in Z} |Z|~ \|H_z\| < \infty
\end{align}

\subsubsection{Matsuta et al. Ref.~\cite{Matsuta_1604}}

In analyzing Ref.~\cite{Matsuta_1604}, we can make use of the results obtained in our iterative procedure after a single iteration.
Using \eqnref{lambda1}, we can immediately compute the exponent of the LC2 power-law light-cone arising from each term of the  bound:
\begin{align}
  \beta^{\mathrm{LC2}}_{n=1;exp}(\sigma) =  \frac{1}{1-\sigma}\, , \, \,\,
  \beta^{\mathrm{LC2}}_{n=1;poly1}(\sigma) = \frac{d+2}{\sigma (\alpha-d) - d} \,
  , \, \,\,
  \beta^{\mathrm{LC2}}_{n=1;poly2}(\sigma) = \frac{2}{\sigma \alpha-d}~.
\end{align}
for $d/(\alpha - d) < \sigma < 1$. This condition immediately requires $\alpha > 2d$ for there to exist a power-law LC2.
Having the exponents as a function of $\sigma$, $\beta^{\mathrm{LC2}}_{\mathrm{M}}$ is given by the optimized exponent with respect to $\sigma$:
\begin{align}
  \beta^{\mathrm{LC2}}_{\mathrm{M}} = \inf_{d/(\alpha-d) < \sigma < 1} \max\left( \beta^{\mathrm{LC2}}_{n=1;\exp}(\sigma), \beta^{\mathrm{LC2}}_{n=1;poly1}(\sigma),\beta^{\mathrm{LC2}}_{n=1;poly2}(\sigma)\right)~.
\end{align}

For all $\sigma$, $\beta^{\mathrm{LC2}}_{n=1;poly1}(\sigma) > \beta^{\mathrm{LC2}}_{n=1;poly2}(\sigma)$ and both are decreasing functions, while $\beta^{\mathrm{LC2}}_{exp}$ is an increasing function. 
As a result, the minima occurs at the intersection between $\beta^{\mathrm{LC2}}_{n=1;poly1}$ and $\beta^{\mathrm{LC2}}_{exp}$, which occurs at $\sigma = (2d+2)/(\alpha+2)$, leading to the light-cone exponent:
\begin{align}\label{eq:LC2_n1}
  \beta^{\mathrm{LC2}}_{n=1} = \frac{\alpha+2}{\alpha - 2d} = \beta^{\mathrm{LC2}}_{\mathrm{M}}~,
\end{align}
for $\alpha > 2d$.

\subsubsection{Our work}

Since we have considered the case of $n=1$ in Appendix D.2.b, we now restrict our attention to $n>1$.
We will begin our calculation by focusing on the contribution from $\beta^{\text{LC2}}_{n;exp}$ and $\beta^{\text{LC2}}_{n;poly2}$ first and then confirming the other polynomial term will not change the obtained exponent.

Based on the exponential term and the polynomial exponents in Eqs.~(\ref{eq:gamma2}) and (\ref{eq:mu2}) we obtain:
\begin{align}
  \beta^{\text{LC2}}_{n;poly2} &= \frac{1+\gamma_2^{(n)}}{-\mu_2^{(n)}-d} = 
  \begin{cases}
    \dfrac{-d + 2\sigma \alpha + n(1+d-\sigma(1+\alpha))}{(1-\sigma)(\sigma\alpha - d)} & \sigma < \dfrac{d+1}{\alpha + 1}\\
    \dfrac{2(1 -\sigma) + d}{(1-\sigma)(\sigma\alpha - d)}& \sigma \ge \dfrac{d+1}{\alpha + 1}
  \end{cases} \label{betaLC2_poly2}\\
  \beta^{\mathrm{LC2}}_{n;exp} &= \frac{1}{1-\sigma}
\end{align}

Because $\beta^{\text{LC2}}_{n;poly2}$ is a convex function, the correct LC2 exponent $\beta^{\text{LC2}}$ will occur in one of two regimes: at the minimum of $\beta^{\text{LC2}}_{n;poly2}$, or at the intersection of $\beta^{\text{LC2}}_{n;poly2}$ and $\beta^{\text{LC2}}_{n;exp}$ (i.e. at the first intersection of
$\beta^{\text{LC2}}_{n;poly2}$ and $\beta^{\text{LC2}}_{n;exp}$; the second intersection is occurs as $\sigma \to 1$, where both exponents become infinite).

The location of the minimum occurs at:
\begin{align}
  &\sigma_{2;min} = 1 + \frac{d}{2} - \frac{d}{2}\sqrt{1 + \frac{2}{d} - \frac{2}{\alpha}} \\
  &\quad \Rightarrow \quad \beta^{\text{LC2}}_{n;poly2}(\sigma_{2;min}) = 2\frac{\alpha-d+d\alpha\left[1 + \sqrt{1 +2/d - 2/\alpha}\right]}{(\alpha- d)^2}
\end{align}

The intersection on the other hand occurs at:
\begin{align}
  \sigma_{2;exp} = \frac{2d+2}{\alpha+2} \quad \Rightarrow \quad \beta^{\text{LC2}}_{n;poly2}(\sigma = \sigma_{2;exp}) = \frac{\alpha+2}{\alpha-2d}
\end{align}
which requires, for consistency, $\alpha > 2d$. This exponent matches that of $n=1$ and Ref.~\cite{Matsuta_1604}.

Because $\sigma_{2;min},\sigma_{2;exp} > (d+1)/(\alpha+1)$, only the second branch of \eqnref{betaLC2_poly2} is relevant for this minimization procedure. This branch is independent of the number of iterations performed, the above results are valid for all $n>1$.

If we now consider $\beta^{\text{LC2}}_{n;poly1}$ it can never improve on this minimization, only worsen it.
Moreover, by choosing $\sigma > (d+1)/(\alpha+1)$ and $n \ge n^*+1$ iterations, one ensures that $\mathfrak{F}_2^{(n*+1)}$ contains the dominant asymptotic time dependence of the polynomial terms, ensuring that $\beta^{\text{LC2}}_{n;poly2} \ge \beta^{\text{LC2}}_{n;poly1}$.
As a result, considering $\beta^{\text{LC2}}_{n;poly1}$ does not change our analysis of the LC2 exponent, only imposes that $n\ge n^*+1$.

Then, by choosing $n\ge n^*+1$, we can immediately compute the LC2 exponent $\beta^{\text{LC2}}$ by just considering $\beta^{\text{LC2}}_{n;poly2}$ and $ \beta^{\text{LC2}}_{n;exp}$.
In this regime, the exponents are independent of $n$, as shown above.
There are two regimes that can determine $\beta^{\text{LC2}}$:
\begin{itemize}
\item $\beta^{\text{LC2}}$ occurs for at the intersection of the curves $\beta^{\text{LC2}}_{n;poly2}(\sigma)$ and $\beta^{\text{LC2}}_{n;exp}$ which occurs at $\sigma = \sigma_{2;exp}$.
  This requires that $\sigma_{2;exp} \le \sigma_{2;min}$, which gives us a condition for $\alpha$:
  \begin{align}
    \alpha \ge \alpha_M \equiv
    \frac{3d}{2} \left(1 + \sqrt{1 + \frac{8}{9d}}\right).
  \end{align}
  which is consistent with the requirement $2d < \alpha_M$ for the intersection solution to be meaningful.
\item $\beta^{\text{LC2}}$ occurs at the minimum of $\beta^{\text{LC2}}_{n;poly2}$ which occurs for $\alpha< \alpha_M$.
\end{itemize}

Thus, we can summarize our result as:
\begin{align}
  \beta^{\text{LC2}} =
  \begin{cases}
    \frac{2}{(\alpha-d)^2}\left\{\alpha-d+d\alpha\left[1 + \sqrt{1 +2/d - 2/\alpha}\right]\right\}& d < \alpha < \alpha_M\\
    \dfrac{\alpha+2}{\alpha-2d} &  \alpha_M < \alpha
  \end{cases}
\end{align}

Then, for $\alpha > \alpha_M$, our LC2 light-cone matches that of Matsuta et al. \cite{Matsuta_1604}.
While for $\alpha<{\alpha_M}$, our iterative procedure ensures a better LC2 under similar assumptions.
In fact, for $\alpha<2d$, our LC2 is well-defined while the LC2 of Matsuta et al. \cite{Matsuta_1604} diverges.
However, our LC2 is bigger than that of a purely
two-body power-law interacting system \cite{FossFeig_1410}.
This situation is summarized in Table \ref{light-cone-table}
and the $\alpha$ dependencies of $\beta^{\mathrm{LC2}}$ are plotted
for the Lieb-Robinson bounds
of the present paper and Refs. \cite{FossFeig_1410,Matsuta_1604}
for $d=1$ in Fig.~\ref{fig:LC2_d1}.


\bibliography{PDTCnew}

\end{document}